\documentstyle[prl,twocolumn,aps,epsfig]{revtex}

\begin{document}

\preprint{CERN-TH/2000-076,nucl-th/0003022}

\title{Tilted Pion Sources from Azimuthally Sensitive HBT Interferometry}

\author{Michael Annan Lisa$^a$, Ulrich Heinz$^b$ and 
        Urs Achim Wiedemann$^b$}
\address{$^a$Department of Physics, The Ohio State University,
174 W. 18th Avenue, Columbus, Ohio 43210\\
$^b$Theoretical Physics Division, CERN, CH-1211 Geneva 23, Switzerland}

\date{\today}

\maketitle

\begin{abstract}
Intensity interferometry in noncentral heavy ion collisions provides 
access to novel information on the geometry of the effective
pion-emitting source. We demonstrate analytically that, even for
vanishing pair momentum, the cross terms $R_{ol}^2$ and $R_{sl}^2$ of
the HBT correlation function in general show a strong first harmonic  
in their azimuthal dependence. The strength of this oscillation 
characterizes the tilt of the major axis of the spatial emission
ellipsoid away from the direction of the beam. Event generator 
studies indicate that this tilt can be large ($>20^\circ$) at AGS
energies which makes it by far the most significant azimuthally
sensitive HBT signal at these energies. Moreover, transport models
suggest that for pions this spatial tilt is directed {\it opposite} 
to the tilt of the directed flow ellipsoid in momentum space. A
measurement of the azimuthal dependence of the HBT cross terms
$R_{ol}^2$ and $R_{sl}^2$ thus probes directly the physical origin of
directed pion flow.\\[2ex] 
\end{abstract} 



Two-particle momentum correlations between
identical particles are commonly used to extract space-time and
dynamical information about the particle emitting source in heavy 
ion collisions. The basis for this intensity interferometric
method is the equation\cite{S73,CH94,WH99} 
 \begin{eqnarray}
   C({\bf q},{\bf K}) &=&
   1 + {\left\vert \int d^4x\, S(x,K)\,
   e^{iq{\cdot}x}\right\vert^2 \over
   \left\vert \int d^4x\, S(x,K)\right\vert^2 } \, ,
 \label{1.1}
 \end{eqnarray}
$q = p_1 - p_2$, $K = \textstyle{1\over 2}(p_1 + p_2)$, 
which relates the phase-space density $S(x,K)$ of the 
source to the measured 2-particle correlation function 
$C({\bf q},{\bf K})$. Experimental measurements of $C$ are 
usually parametrized in terms of the intercept $\lambda({\bf K})$ 
and the HBT radii $R_{ij}^2({\bf K})$ by
\begin{equation}
  \label{1.2}
    C({\bf q},{\bf K}) =
    1 + \lambda({\bf K})\, 
    \exp\Bigl[- \sum_{i,j=o,s,l} q_i q_j R_{ij}^2({\bf K}) \Bigr]
    \,.
\end{equation}
In this Cartesian {\it osl}-system the relative momentum is
decomposed into components parallel to the beam ($l$ = {\it longitudinal}), 
parallel to the transverse component of ${\bf K}$ ($o$ = {\it out}), 
and in the remaining third direction ($s$ = {\it side}). 

Over the last decade the experimental frontier in studying identical
two-particle correlations was defined by more and more differential
measurements of the HBT radii $R_{ij}^2({\bf K})$. At both AGS
\cite{E877,E895} and SPS \cite{NA49} energies, the longitudinal
($K_L$) and transverse ($K_\perp$) pair momentum dependence is now
well-studied for central and reaction-plane averaged non-central
collisions. Most importantly, these studies have led to a detailed
characterization of the longitudinal expansion and the transverse
radial flow of the reaction zone. The next challenge is a similarly
detailed study of the $R_{ij}^2({\bf K})$ as a function of the
azimuthal orientation $\Phi$ of the transverse pair momentum 
${\bf K}_\perp$ with respect to the impact parameter ${\bf b}$ in
non-central collisions. This $\Phi$-dependence reveals qualitatively 
new infor\-ma\-tion about the space-time structure of the source and 
yields new insights on the underlying nature of flow. In fact, under
reasonable assumptions {\em all 10 components of the source's spatial
  correlation tensor can be recovered.} 

An azimuthally sensitive HBT analysis involves all six parameters 
$R_{ij}^2$, all of which are functions of all 3 components of the 
pair momentum $K_\perp$, $Y$ and $\Phi$ \cite{fn1}. They provide
information about the source in space-time according to the following
relations~\cite{W98,WH99}: 
  \begin{eqnarray}
    && R_s^2(K_\perp,\Phi,Y) = S_{11} \sin^2\Phi
                  + S_{22} \cos^2\Phi
                  - S_{12} \sin 2\Phi \, ,
                      \nonumber \\
    && R_o^2(K_\perp,\Phi,Y) = S_{11} \cos^2\Phi 
                  + S_{22} \sin^2\Phi
                  + S_{12} \sin 2\Phi
   \nonumber \\
    && \qquad \qquad \quad 
                     - 2\beta_\perp S_{01} \cos\Phi 
                     - 2\beta_\perp S_{02} \sin\Phi 
                     + \beta_\perp^2 S_{00} \, ,
   \nonumber \\
    && R_{os}^2(K_\perp,\Phi,Y) = 
                  S_{12} \cos 2\Phi 
                  + \textstyle{1\over 2} \left(S_{22}-S_{11}\right)
                  \sin 2\Phi 
   \nonumber \\
    && \qquad \qquad \qquad \ \ 
                     + \beta_\perp S_{01} \sin\Phi
                     - \beta_\perp S_{02} \cos\Phi \, ,
   \nonumber \\
    && R_{l}^2(K_\perp,\Phi,Y) = S_{33} -2 \beta_l S_{03} +
                       \beta_l^2 S_{00} \, ,
   \nonumber \\
    && R_{ol}^2(K_\perp,\Phi,Y) = 
                  \left( S_{13} - \beta_l S_{01}\right) \cos\Phi
                  - \beta_\perp S_{03} 
   \nonumber \\
    &&\qquad \qquad \qquad \ 
                  + \left( S_{23} - \beta_l S_{02}\right) \sin\Phi
                  + \beta_l\beta_\perp S_{00}\, ,
   \nonumber \\
    && R_{sl}^2(K_\perp,\Phi,Y) = 
                  \left(S_{23} - \beta_l S_{02}\right) \cos\Phi
   \nonumber \\
    &&\qquad \qquad \qquad \ 
                  - \left( S_{13} - \beta_l S_{01}\right) \sin\Phi\, .
    \label{1.3}
  \end{eqnarray}
The pair velocity $\bbox{\beta}{\,=\,}{\bf K}/K^0$ arises from the 
on-shell constraint \cite{WH99} $q^0{\,=\,}{\bf q}{\cdot}\bbox{\beta}$; 
it mixes spatial and temporal information. $S_{\mu\nu}$ denotes the 
spatial correlation tensor
\begin{equation}
  S_{\mu\nu} = \langle \tilde x_\mu \tilde x_\nu \rangle\, ,
  \quad \tilde{x}_\mu = x_\mu - \bar x_\mu\, ,
  \quad (\mu,\nu=0,1,2,3)
  \label{1.4}
\end{equation}
which measures the Gaussian width in space-time of the emission function
$S(x,K)$ around the point of highest emissivity 
$\bar{x}_\mu = \langle \tilde x_\mu \rangle$ \cite{WH99}:
\begin{eqnarray}
   \langle \tilde x_\mu \tilde x_\nu \rangle (K) &=& 
   {\int d^4x\, \tilde x_\mu \tilde x_\nu\, S(x,K)
     \over\int d^4x \, S(x,K)} \, .
    \label{1.5} 
\end{eqnarray}
It is the inverse of the curvature tensor $B_{\mu\nu}$ \cite{CNH95}:
\begin{equation}
  S(x,K) \approx N(K)\, S(\bar{x},K)\, 
  \exp\Bigl[ -{\textstyle{1\over 2}} \tilde{x}^\mu
              B_{\mu\nu} \tilde{x}^\nu\Bigr]\, .
  \label{1.6}
\end{equation}
This approximation neglects non-Gaussian components of the emission 
function whose influence on the HBT radii can in most practical
cases be neglected~\cite{WH99}. We emphasize that in (\ref{1.4}) 
$S_{\mu\nu}$ is defined in terms of Cartesian coordinates in an 
impact parameter fixed system, in which $x_1=x$ is parallel to the 
impact parameter ${\bf b}$ and $x_3=z$ lies in the beam direction.

The general relations (\ref{1.3}) separate the {\it explicit} 
$\Phi$-dependence of the HBT-radii (which is a consequence of the 
azimuthal rotation of the $osl$-system relative to $(x_1,x_2,x_3)$) 
from the {\it implicit} $\Phi$-dependence of the space-time widths 
$\langle \tilde{x}_{\mu}\tilde{x}_{\nu}\rangle(K_\perp,Y,\Phi)$
(which reflects a $\Phi$-dependent change of the shape of the
effective emission region)~\cite{W98}. Existing studies of (\ref{1.3})
focussed on the detailed interplay between explicit and implicit
$\Phi$-de\-pen\-dences in the HBT radii $R_s^2$, $R_o^2$ and
$R_{os}^2$ \cite{W98,VC96,H99}. Here we show, however, that some of
the most striking features are found in analyzing the
$\Phi$-dependences of $R_{ol}^2$ and $R_{sl}^2$ which so far received
less attention \cite{VC96}. 

The following discussion is simplified significantly by the
important observation that the implicit $\Phi$-de\-pen\-dence of 
$S_{\mu\nu}$ is weak. It can be neglected relative to the explicit 
one given in (\ref{1.3}) as long as the $\Phi$-dependence of
space-momentum correlations in the source is small compared to the 
thermal smearing, and for $K_\perp \to 0$ it vanishes completely.
Studies with the RQMD model \cite{rqmd} indicate that the first 
condition works well at least up to $p_T = 300$ MeV/c for Au+Au 
collisions at 2 $A$\,GeV~\cite{LHW00}. Beyond such model studies,
a simple scale argument illustrates why neglecting the implicit
$\Phi$-dependence relative to the explicit one has a much wider
kinematical region of validity than neglecting the implicit 
$K_\perp$-dependence relative to the explicit one in (\ref{1.3}):
the latter is suppressed near $K_\perp{\,=\,}0$ ($\beta_\perp{\,=\,}0$), 
and it multiplies only space-time variances involving $\tilde{t}$ which 
are numerically small in practice. In contrast, the explicit
$\Phi$-dependence in (\ref{1.3}) leads to prefactors $\cos(n\Phi)$,
$\sin(n\Phi)$ oscillating between $1$ and $-1$ even for $K_\perp = 0$
, and it multiplies the numerically large components of
$S_{\mu\nu}$. The assumption of vanishing implicit $\Phi$-dependence
can be checked experimentally~\cite{W98}, and deviations can be
quantified in a full harmonic analysis given elsewhere
\cite{W98,LHW00}. Also, while it requires weak {\it transverse} flow, 
there is no such restriction on the longitudinal flow. Qualitatively,
the main findings presented here do not depend on this assumption, but
it simplifies our presentation and allows for a particularly intuitive
geometric picture of the new effect discussed here.  

With this proviso, the components $S_{\mu\nu}$ in (\ref{1.3}) 
become $\Phi$-independent constants which describe the {\it same} 
source being viewed from all angles $\Phi$. We turn briefly to 
symmetry considerations at midrapidity. Considering collisions between 
equal mass nuclei, it can be rigorously shown \cite{LHW00} that, 
as a consequence of point reflection symmetry around the spatial
origin and mirror symmetry with respect to the reaction plane, 
five of the off-diagonal components $S_{\mu\nu}$ (all except 
$S_{13}$) oscillate symmetrically around zero.  Coupled with the
condition of vanishing implicit $\Phi$-dependence this implies
\begin{equation}
  S_{01} = 0\, ,\;
  S_{02} = 0\, ,\;
  S_{03} = 0\, ,\;
  S_{12} = 0\, ,\;
  S_{23} = 0\, .
  \label{1.7}
\end{equation} 
These equations and the fact that around midrapidity the average
$\beta_l$ is zero (although average $\beta_l^2 \neq 0$) allow us to
write the HBT radius parameters (\ref{1.3}) in terms of 5
non-vanishing components only: 
  \begin{eqnarray}
    && R_s^2 = \textstyle{1\over 2} \left(S_{11} + S_{22}\right)
             + \textstyle{1\over 2} \left(S_{22} - S_{11}\right)
               \cos 2\Phi\, ,
  \nonumber \\
    && R_o^2 = \textstyle{1\over 2} \left(S_{11} + S_{22}\right)
             - \textstyle{1\over 2} \left(S_{22} - S_{11}\right) \cos 2\Phi\,
             + \beta_\perp^2 S_{00}\, ,
  \nonumber \\
    && R_{os}^2 = \textstyle{1\over 2} \left(S_{22}-S_{11}\right) \sin 2\Phi 
  \nonumber \\
    && R_{l}^2 = S_{33} + \beta_l^2 S_{00} \, ,
  \nonumber \\
    && R_{ol}^2 = S_{13}\cos\Phi\, , 
  \nonumber \\
    && R_{sl}^2 = - S_{13} \sin\Phi\, .
  \label{1.8}
  \end{eqnarray}
Since the $\Phi$-dependences of $R_s^2$ and $R_o^2$ explicitly
separate $S_{22}$ from $S_{11}$, the emission duration $S_{00}$ can
now be determined without additional model assumptions \cite{WH99,H99}, 
contrary to the case of collisions at $b{\,=\,}0$. 

Given the measured weak $Y$-dependence of the HBT-radii 
\cite{WH99,E877,E895,NA49}, Eqs.~(\ref{1.8}) can be used in practice
also for event samples which are averaged over large $Y$-windows
symmetric around $Y=0$. According to (\ref{1.8}), the HBT radius
parameters $R_{o}^2$, $R_{s}^2$ and $R_{os}^2$ all show second
harmonic oscillations of the same strength
$\frac{1}{2}\left(S_{11}{-}S_{22}\right)$. This is the
${R_{o,2}^c}^2{\,=\,}-{R_{s,2}^c}^2{\,=\,}-{R_{os,2}^s}^2$ rule for
second harmonic coef\-ficients~\cite{W98}; leading deviations from
this rule have been quantified~\cite{W98,LHW00} and provide a
consistency check on the assumption of negligible implicit
$\Phi$-dependence. More strikingly, $R_{ol}^2$ and $R_{sl}^2$ display
purely {\it first harmonic oscillation at midrapidity} which are
easier to measure. The expected identical amplitudes for these
oscillations provide a further consistency check on our assumptions. 

The required $\Phi$-binning puts severe demands on the pair 
statistics. A first observation of azi\-mu\-thally oscillating HBT 
radii was reported in \cite{E895_QM99}. Such measurements require a 
reasonably accurate determination of the reaction plane; a typical 
uncertainty of 30$^\circ$ reduces the first and second 
harmo\-nics in (\ref{1.3}) by $\sim 15$\% and $\sim 45$\%, 
respectively, but these losses can be corrected for \cite{W98,VZ96}.

While the amplitude of the oscillations of $R_{o}^2$, $R_{s}^2$, 
and $R_{os}^2$ are given by the difference between the transverse
source sizes in and perpendicular to the reaction plane, that of the
oscillations of $R_{sl}^2$ and $R_{ol}^2$ is given by $S_{13} \equiv
\langle \tilde x\, \tilde z\rangle$. Parameterizing the source by an
ellipsoid, a nonzero $S_{13}$ corresponds to a tilt in the reaction
plane of the longitudinal major axis of the ellipsoid away from the
beam direction. It can be characterized by a tilt angle 
\begin{equation}
  \theta_s = \frac{1}{2} \tan^{-1}
       \left( \frac{2S_{13}}{S_{33}-S_{11}}\right)\, .
  \label{1.9}
\end{equation}
Rotating the spatial correlation tensor $S_{\mu\nu}$ by 
$\theta_s$ yields a purely diagonal tensor 
$S^{\prime} = R_y^\dagger (\theta_s) \cdot S \cdot R_y(\theta_s)$
whose eigenvalues are the squared lengths of the 3 major axes.

\begin{figure}
\vspace*{-0.9cm}
\begin{center}
\epsfig{file=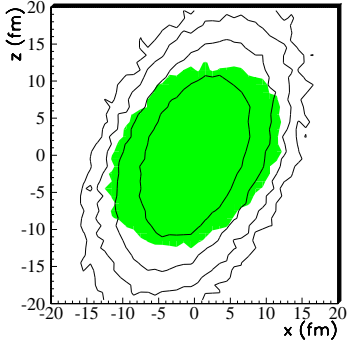,width=5cm}
\end{center}
\vspace*{-0.2cm}
\caption{
The tilt of the spatial distribution of pion emission points 
projected onto the reaction ($xz$) plane. Contours show on a 
logarithmic scale the spatial distribution for the toy model 
(\protect\ref{1.10}) with $\sigma_{t}\,{=}\,5$\,fm/$c$,  
$\sigma_x\,{=}\,4$\,fm, $\sigma_y\,{=}\,5$\,fm, $\sigma_z\,{=}\,7$\,fm, 
and $\Theta = 25^\circ$. The shaded region is the distribution of 
emission points of pions with $|p_z|<40$ MeV/c in the toy 
model including longitudinal flow. See text for details.
\label{fig1}}
\end{figure}

We illustrate the role of the tilt angle (\ref{1.9}) with a tilted 
Gaussian toy distribution with no space-momentum correlations:
\begin{eqnarray}
  && S(x,K) =  e^{-E/T} \exp\left(
                       -\frac{x^{\prime 2}}{2\sigma_{x}^2}
                       -\frac{y^2}{2\sigma_y^2}
                       -\frac{z^{\prime 2}}{2\sigma_{z}^2}
                       -\frac{t^2}{2\sigma_t^2}
                               \right) \,,
  \nonumber \\
  && x^\prime = x\cos\Theta - z\sin\Theta\, , \quad 
     z^\prime = x\sin\Theta + z\cos\Theta .
  \label{1.10}
\end{eqnarray}
To avoid relativistic complications, the ``temperature'' $T$ is kept 
small (20 MeV) in the following. Fig.~\ref{fig1} shows the projection 
of this source onto the reaction ($xz$) plane.

\begin{figure}
\vspace*{-1.7cm}
\begin{center}
\epsfig{file=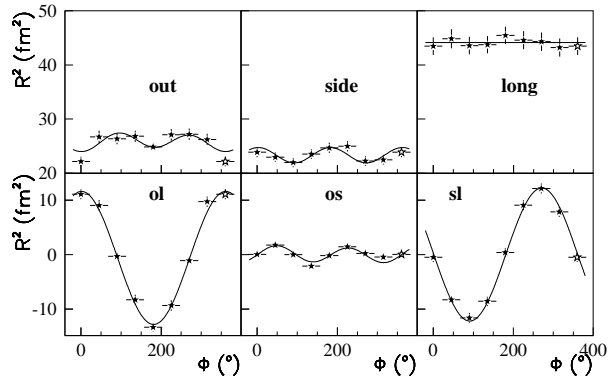,width=\linewidth}
\end{center}
\vspace*{-.1cm}
\caption{
Azimuthal dependence of the HBT radii from fits to correlation functions 
generated from the toy source (\protect\ref{1.10}). Solid lines represent 
a global fit to $R(\Phi)$ with (\protect\ref{1.3}). The value from 
$\Phi=0^\circ$ is replotted with an open symbol at $\Phi=360^\circ$.
\label{fig2}}
\end{figure}

%
Using the model (\ref{1.10}) with the parameters of Fig.~\ref{fig1}
to randomly generate a set of phase-space points, we constructed
a three-dimensional correlation function (with $\approx 4 \times 10^5$ 
pairs with $q<100$ MeV/c) for each of eight $45^\circ$-wide $\Phi$ bins, 
according to the prescription and code of Pratt \cite{Pratt}. Fitting 
each with the Gaussian parametrization (\ref{1.2}) yields the HBT 
radii presented in Fig.~\ref{fig2}. Treating the ten components of 
$S_{\mu\nu}$ as parameters, we perform a global fit with Eqs.~(\ref{1.3}) 
on these $\Phi$-dependent radii. The fit results are indicated by solid 
lines in Fig.~\ref{fig2}. Application of (\ref{1.9}) to the fit results 
yields $\theta_s=24.6^\circ\pm 0.6^\circ$, in good agreement with the
input value $\Theta{\,=\,}25^\circ$. The diagonal elements 
$S^\prime_{\mu\mu}$ reproduce, within statistical errors, the input 
values for the squares of the homogeneity lengths:
$\sigma_x = 4.05 \pm 0.05$ fm,
$\sigma_y = 4.97 \pm 0.04$ fm,
$\sigma_z = 6.97 \pm 0.04$ fm, and
$\sigma_t = 4.78 \pm 0.45$ fm. (The larger uncertainty in
$\sigma_t$ arises from the low $\beta_\perp$ of the pions in
our example.)

While one may escape the effects of transverse flow (which may generate 
a $\Phi$-dependent effective source) by selecting pion pairs at
low $K_\perp$, longitudinal flow, which generates $z-p_z$ correlations, 
is generally stronger and cannot be cut away. Fortunately, since they
are essentially orthogonal to the azimuthal dependences we are discussing, 
such correlations do not drastically alter the intuitive geometric picture 
we have discussed -- the same source is still viewed from all angles $\Phi$.

As an example we added a boost-invariant longitudinal flow component 
in $z$-direction to our toy source (scaled so that the collective 
flow velocity at $z= \pm \sigma_z$ is equal to the thermal velocity),
leaving the geometry unchanged. This results in (i) an increase in the 
tilt angle $\theta_s$ from 25$^\circ$ to 33$^\circ$ and (ii) a 
reduction in $S^{\prime}_{33}$ from 49 fm$^2$ to 31 fm$^2$.  
The other components $S^{\prime}_{\mu\mu}$ vary negligibly from the 
scenario without flow. Familiar from the case of azimuthally symmetric 
HBT, effect (ii) is understood in terms of a reduction in the length of 
homogeneity due to the flow~\cite{WH99}: HBT correlations arise from
particle pairs with close-by momenta; the space-momentum correlations 
induced by longitudinal flow then imply that they will be close-by
in coordinate space as well. The increased tilt is similarly understood, 
by examining the spatial distribution of emission points for pions with 
low $p_z$. The shaded region in Fig.~\ref{fig1} shows the effective 
source for pions with $|p_z| < 40$ MeV/$c$; it is clearly less prolate 
and more tilted than in the case of no flow (contour lines).

\vspace{0.3cm}

\phantom{n}
%
%
$\Phi$-sensitive HBT studies and the measurement of spatial tilt
connect for the first time the physics of directed flow with the 
space-time structure of the source \cite{VC96}. We con\-si\-der 
this an essential step towards a full understanding of the 
phase-space dynamics of heavy-ion collisions. Although a detailed 
discussion must await a longer paper, we here shortly touch on the 
main physics points, using results from a realistic transport model. 

\begin{figure}
\vspace*{-0.9cm}
\begin{center}
\epsfig{file=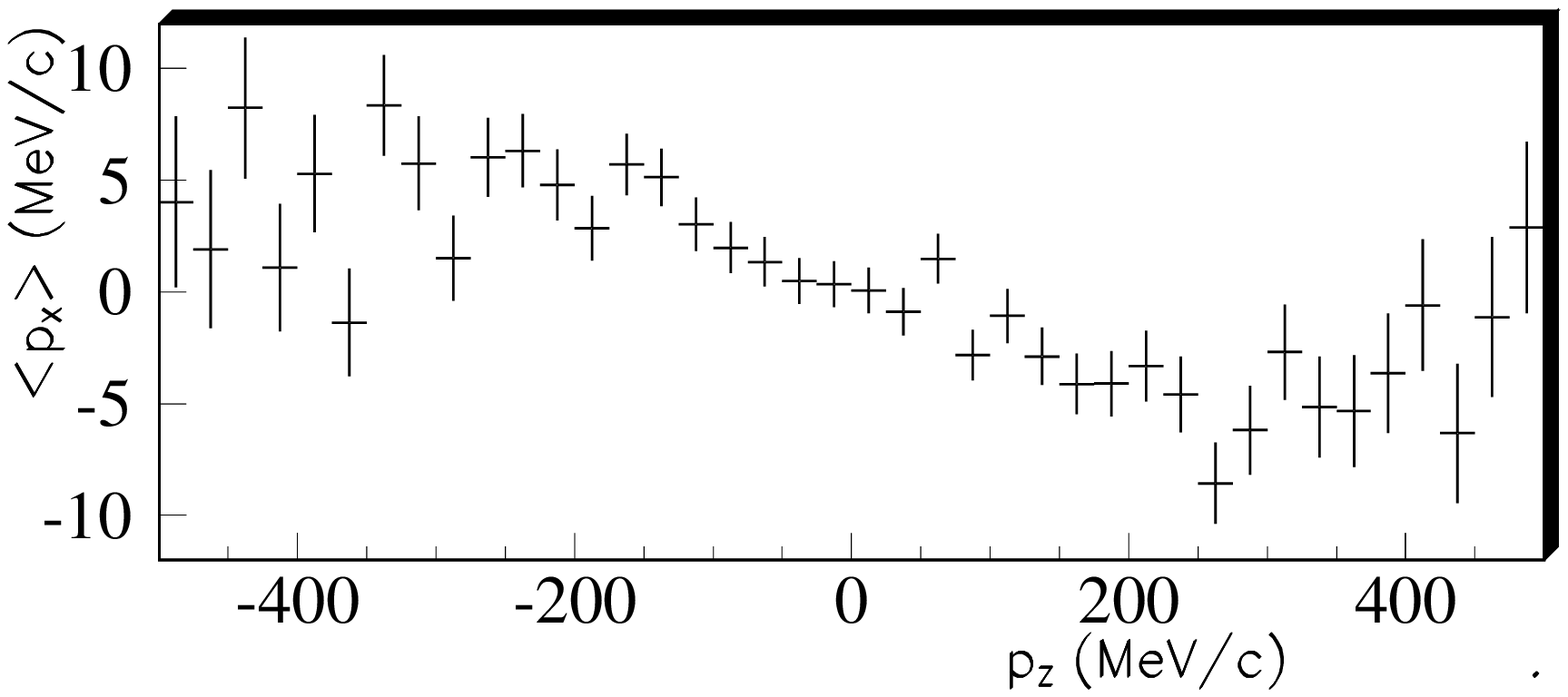,width=6cm}
\epsfig{file=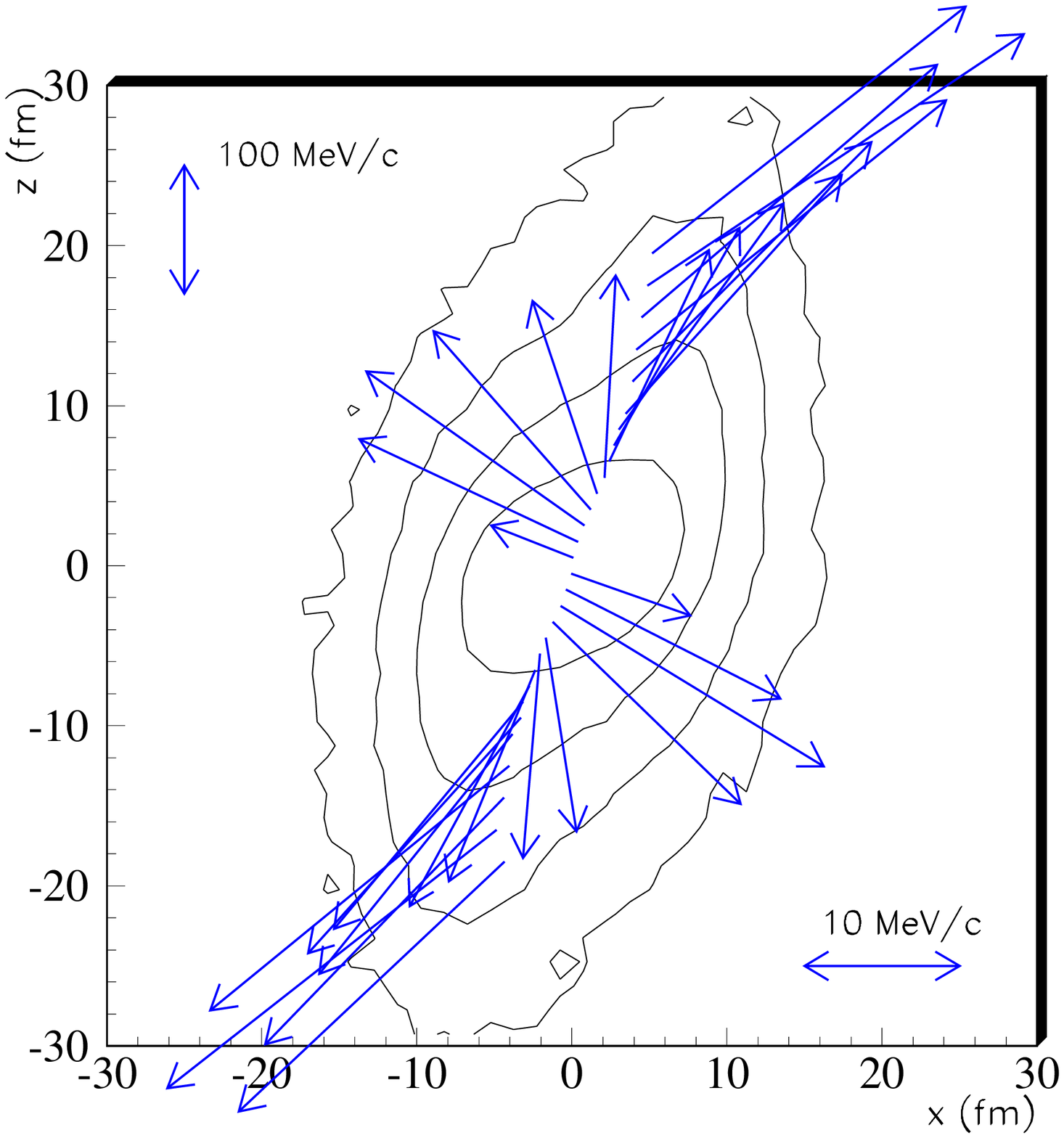,width=6cm}
\end{center}
\caption{
RQMD simulation of pions from 2 $A$\,GeV Au+Au collisions at
$b$=3-7\,fm. The top panel shows a weak $p_x{-}p_z$ ``anti-flow'' 
correlation. In the bottom panel, contours of the spatial distribution 
of emission points projected onto the reaction plane show a strong tilt 
in the {\it opposite direction from the tilt in momentum space}. 
Superimposed arrows represent the average pion momentum at different 
values of $z$. Note that the momentum scale in $z$-direction is
compressed for clarity.
\label{fig3}}
\end{figure}

We performed simulations of semiperipheral Au+Au collisions at 
2 $A$\,GeV with the RQMD (v2.3) model~\cite{rqmd}. The top panel of 
Fig.~\ref{fig3} shows $\langle p_x \rangle$ -- the average pion 
momentum in the reaction plane -- as a function of momentum $p_z$
along the beam axis. Qualitatively consistent with experimental
observations~\cite{EOS_pionflow}, a very weak negative directed 
flow (``anti-flow'') signal is observed -- the average emission 
ellipsoid in {\it momentum} space is tilted to a negative angle with 
respect to the beam (the direction of directed proton flow defines 
the positive direction). The magnitude of the collective motion 
($\sim 10$\,MeV/$c$) is small compared to the typical $p_T$ scale 
($\sim 200$\,MeV/$c$); hence thermal smearing dominates. The 
bottom panel shows that, while the spatial distribution displays 
a richer structure than our toy model, it is nevertheless always 
characterized by a significant {\it positive} tilt -- {\it opposite} 
the average tilt in momentum space. A full correlation function
ana\-ly\-sis of the RQMD events \cite{LHW00} yields qualitatively
similar results as those shown in Fig.~\ref{fig2}. 

%
%

This bears directly on the physical causes of directed pion flow at
these energies. Detailed transport model studies \cite{bass} have
shown that pion reflection from (not absorption by) the nucleonic
matter is at the root of directed pion flow at these energies. 
Focussing on the forward hemisphere, if absorption processes 
($\pi NN \to \Delta N \to NN$) were dominant in producing pion
flow, we would expect an absence of $\pi$ emission points in the 
$+x$ quadrant, i.e. a negative tilt in coordinate space {\it and} 
in momentum space. Since it is the point of last scattering (as
opposed to the original point of creation) which is relevant for HBT
correlations \cite{WH99}, it is clear that reflection ($\pi N \to
\Delta \to \pi N$) from flowing participant or spectator baryons leads
to a positive tilt in coordinate space as seen in Fig.~\ref{fig3}: the
reflected pions ``illuminate'' the coordinate-space anisotropies of
the nucleonic matter. In this simple picture, then, the sign of 
$\theta_s$ immediately distinguishes between these two possibilities. 

The arrows in Fig.~\ref{fig3} represent the average momenta of pions 
for different values of $z$. The resulting structure further 
underscores the importance of pion rescattering: Clearly, the more 
numerous pions from the high-density region around $z=0$ dominate, 
generating the anti-flow signal seen in experiment.
However, pions from the more dilute large-$|z|$ region have less 
opportunity for rescattering and so retain the {\it positive} 
$p_x{-}p_z$ correlation of their (flowing) parent $\Delta$'s.  
Similar considerations generate a sign change in the pion flow 
as the impact parameter is varied in transport models~\cite{bass}.

In summary, for non-central collisions all ten components of the 
spatial correlation tensor $S_{\mu\nu}$ are accessible by 
$\Phi$-dependent HBT measurements. Based on symmetry and scale 
considerations we argue that for low $K_\perp$ the explicit 
$\Phi$-dependence of Eqs.~(\ref{1.3}) dominates. Consistency relations
allow to check whether this is true in practice. The spatial
correlation tensor $S_{\mu\nu}$ can then be extracted completely from
a global fit to the six $\Phi$-de\-pen\-dent HBT radii. At
midrapidity, the five nonvanishing components of $S_{\mu\nu}$  
correspond to the four spacetime lengths of homogeneity and a tilt 
of the source in the reaction plane, away from the beam direction. 
This tilt, which may be quite large at AGS energies, causes striking 
and relatively easily measurable first-order harmonic oscillations in 
$R_{ol}$ and $R_{sl}$ and can give a direct experimental handle on 
the origin of pion flow at these energies.

The work of M.A.L. is supported by NSF Grant PHY-9722653 and that of
U.H. by DFG, GSI and BMBF.


\end{document}